\documentclass[12pt]{iopart}
\usepackage{iopams,mathptm,times,graphicx}

\makeatletter

\newcounter{theorem}
\@addtoreset{theorem}{section}
\renewcommand\thetheorem{\arabic{section}.\arabic{theorem}}

\newenvironment{proof}{\par\noindent{\bf Proof.} }{\proofbox\par\medskip}

\def\proofbox{\hfill{\ensuremath\Box}}

\newdimen\LENB \newdimen\LENW \newdimen\THI
\newdimen\LENWH \newdimen\LENTOT \newcount\N
\def\vbrknlnele#1#2#3{
  \LENB=#1pt \LENW=#2pt \THI=#3pt
  \LENWH=\LENW \divide\LENWH by 2
  \LENTOT=\LENB \advance\LENTOT by \LENW
  \vbox to \LENTOT{
    \vbox to \LENWH{}
    \nointerlineskip
    \vbox to \LENB{\hbox to \THI{\vrule width \THI height \LENB}}
    \nointerlineskip
    \vbox to \LENWH{}
  }}

\def\vbrknln#1{
  \N=#1
  \vcenter{
    \vbox{
      \loop\ifnum\N>0
        \vbox to 4pt{\vbrknlnele{2}{2}{0.1}}
        \nointerlineskip
        \advance\N by -1
      \repeat
  }}}

\def\hbrknlnele#1#2#3{
  \LENB=#1pt \LENW=#2pt \THI=#3pt
  \LENTOT=\LENB \advance\LENTOT by \LENW
  \vcenter{
    \vbox to \THI{
      \hbox to \LENTOT{
        \hfil
        \vrule width \LENB height \THI
        \hfil}
  }}}

\def\img{{\rm i}}

\makeatletter

\usepackage{ulem}

\def\journal#1&#2,{\begingroup \let\journal=\dummyjournal
               \it #1\unskip~\bf\ignorespaces #2\rm,\endgroup}
\def\dummyjournal{\errmessage{Reference foul up: nested \journal macros}}

\eqnobysec

\def\eqref#1{(\ref{#1})}

\begin{document}
\title[Integrable discretizations of the short pulse equation]
  {Integrable discretizations of the short pulse equation}
\author{Bao-Feng Feng$^1$, Ken-ichi Maruno
$^1\footnote{Corresponding author; e-mail: kmaruno@utpa.edu}$ and 
Yasuhiro Ohta$^{2}$ 
}
\address{$^1$~Department of Mathematics,
The University of Texas-Pan American,
Edinburg, TX 78541
}
\address{$^2$~Department of Mathematics,
Kobe University, Rokko, Kobe 657-8501, Japan
}
\date{\today}
\def\submitto#1{\vspace{28pt plus 10pt minus 18pt}
     \noindent{\small\rm To be submitted to : {\it #1}\par}}

\begin{abstract}
In the present paper, we propose integrable semi-discrete and
full-discrete analogues of the short pulse (SP) equation.
The key of the construction is the bilinear forms and
determinant structure of solutions of the SP equation.
We also give the determinant formulas of $N$-soliton solutions of the
semi-discrete and full-discrete analogues of the SP equations, from
which the multi-loop and multi-breather solutions can be generated.
In the continuous limit, the full-discrete SP equation
converges to the semi-discrete SP equation, then to the continuous SP
equation. Based on the semi-discrete SP equation, an integrable
numerical scheme, i.e., a self-adaptive moving mesh scheme, is
proposed and used for the numerical computation of the short pulse
equation.
\par
\kern\bigskipamount\noindent
\today
\end{abstract}

\kern-\bigskipamount
\pacs{02.30.Ik, 05.45.Yv, 42.65.Tg, 42.81.Dp}

\submitto{\JPA}

\section{Introduction}
Most recently, the short pulse (SP) equation
\begin{equation}
u_{xt}=u+\frac{1}{6}(u^3)_{xx}
\end{equation}
was derived as a model equation for the
propagation of ultra-short optical pulses in nonlinear
media \cite{SP,SP2}.
Here, $u=u(x,t)$ represents the magnitude of the electric field,
the subscripts $t$ and $x$ denote partial differentiation. 
Apart from the context of nonlinear optics,
the SP equation has also been derived
as an integrable differential equation associated
with pseudospherical surfaces \cite{Robelo}.
The SP equation has been
shown to be completely integrable
\cite{Robelo,Beals,Sakovich,Brunelli1,Brunelli2}.
The loop soliton solutions as well as
smooth soliton solutions of the SP equation were found in 
\cite{Sakovich2,Kuetche,Kuetche2}.
The connection between the SP equation and the sine-Gordon 
equation through the hodograph transformation
was clarified in \cite{Matsuno},
and then the $N$-soliton solutions including multi-loop
and multi-breather ones were given by using the Hirota bilinear method. 

Integrable discretizations of soliton equations
have received considerable attention recently \cite{SIDE,DIS,Suris,Bobenko}.
In our recent work, the authors proposed an integrable semi-discrete
analogue of the Camassa-Holm (CH) equation and
apply it as a numerical scheme, i.e., 
a self-adaptive moving mesh scheme \cite{dCH,dCH2}.
The key of the discretization is an introduction of an nonuniform mesh, 
which plays a role of the hodograph
transformation as in the continuous case.

In the present paper, we attempt to construct integrable semi-discrete and
full-discretizations of the SP equation by the same 
approach used in the CH equation.
We also attempt to use the semi-discrete analogue of the SP equation as a self-adaptive moving mesh scheme to perform
numerical simulations.

The rest of the present paper is organized as follows. In Section 2, we
review the bilinear equations and determinant solutions of the SP
equation. In Section 3, we propose an integrable
semi-discrete analogue of the SP equation, whose $N$-soliton solutions are also constructed in terms of determinant form. By using the semi-discrete
analogue of the SP equation as a self-adaptive moving mesh scheme, 
the numerical results for one- and two-loop solutions are also presented.
In Section 4, the full-discrete analogues of the SP equation are
proposed. The paper is concluded by Section 5. 

\section{Bilinear equations and 
determinant solutions of the short pulse equation}

In this section, the results in \cite{Matsuno} regarding the bilinear equations
and the solutions of the SP equation will be briefly reviewed.

First, by introducing the new dependent variable
\begin{equation}
r^2=1+u_x^2 \,,
\end{equation}
the SP equation is rewritten as
\begin{equation}
r_t=\left(\frac{1}{2}u^2r\right)_x\,.
\end{equation}
Introducing the hodograph transformation
\begin{equation}
dy=rdx+\frac{1}{2}u^2rdt\,,\quad ds=dt\,,
\end{equation}
i.e.,
\[
\frac{\partial}{\partial t}=\frac{1}{2}u^2r\frac{\partial}{\partial y}
+ \frac{\partial}{\partial s}\,,\quad
\frac{\partial}{\partial x}=r\frac{\partial}{\partial y}\,,
\]
we obtain
\begin{equation}
r_s=r^2uu_y\,,\label{r-u-eq}
\end{equation}
where 
\[
 r^2=1+r^2u_y^2\,.
\]
The equation (\ref{r-u-eq}) can also be cast into
a form of
\begin{equation}
 \left(\frac{1}{r}\right)_s=-\left(\frac{1}{2}u^2\right)_y\,.
 \label{tmp1}
\end{equation}
Introducing new variables
\begin{equation}
r=\frac{1}{\cos \phi}\,,\quad u=\phi_s\,,
\end{equation}
Eq.(\ref{tmp1}) leads to the sine-Gordon equation
\begin{equation}
\phi_{ys}=\sin \phi\,.
\label{SG}
\end{equation}

Moreover, as is shown in \cite{Hirota,HirotaBook},
upon the dependent variable transformation
\[
 \phi(y,s)=2\img\, \ln \frac{F^*(y,s)}{F(y,s)}\,,
\]
the sine-Gordon equation (\ref{SG}) leads to 
the following bilinear
equations
\begin{eqnarray}
&&FF_{ys}-F_yF_s=\frac{1}{4}(F^2-{F^*}^2)\,, \label{bilinear_SG1}\\
&&F^*F_{ys}^*-F_y^*F_s^*=\frac{1}{4}({F^*}^2-F^2)\,,
\label{bilinear_SG2}
\end{eqnarray}
where $F^*$ is the complex conjugate of $F$.
Henceforth, the solutions of the SP equation are obtained 
by $F$ and $F^*$ through the dependent variable transformation 
\begin{equation}
u(y,s)
=\frac{\partial}{\partial s}
\phi(y,s)=\frac{\partial}{\partial s}\left(2\img \, \ln
\frac{F^*(y,s)}{F(y,s)}\right)\,.
\end{equation}

In what follows, we will show that the bilinear 
equations (\ref{bilinear_SG1})--(\ref{bilinear_SG2}) are 
actually obtained as 
the 2-reduction of the two-dimensional Toda lattice (2DTL) equations:
\cite{Mikhailov,Hirota-Toda,HIK,HirotaBook}
\begin{equation}
\frac{1}{2}D_YD_S \tau_n\cdot \tau_n
={\tau_n}^2-{\tau_{n+1}}{\tau_{n-1}}\,,
\end{equation}
i.e.,
\begin{equation}
\tau_n\frac{\partial^2{\tau_n}}{\partial Y\partial S}-
\frac{\partial {\tau_n}}{\partial Y}\frac{\partial {\tau_n}}{\partial S}
={\tau_n}^2-{\tau_{n+1}}{\tau_{n-1}}\,,
\end{equation}
where $D_x$ is the Hirota $D$-operator which is
defined as
\[
D_x^nf\cdot g=\left(\frac{\partial}{\partial x}
-\frac{\partial}{\partial y}\right)^nf(x)g(y)|_{y=x}\,.
\]
Applying the 2-reduction
$\tau_{n-1}=\alpha^{-1} \tau_{n+1}$ ($\alpha$ is a constant),
we obtain
\begin{equation}
\tau_n\frac{\partial^2{\tau_n}}{\partial Y\partial S}-
\frac{\partial {\tau_n}}{\partial Y}\frac{\partial {\tau_n}}{\partial S}
={\tau_n}^2-\tau_{n+1}^2\,,
\end{equation}
where the gauge transformation
$\tau_n \to \alpha^{\frac{n}{2}} \tau_n$
is used.
Letting $f=\tau_{0}$ and $\bar{f}=\tau_{1}$,
we have
\begin{eqnarray}
&&ff_{YS}-f_Yf_S=f^2-{\bar{f}}^2\,,\\
&&\bar{f}\bar{f}_{YS}-\bar{f}_Y\bar{f}_S={\bar{f}}^2-f^2\,.
\end{eqnarray}
Under the independent variable transformation
$y=2Y$, $s=2S$,
we obtain
\begin{eqnarray}
&&ff_{ys}-f_{y}f_{s}=\frac{1}{4}(f^2-{\bar{f}}^2)\,,\label{bilinear-SP1}\\
&&\bar{f}\bar{f}_{ys}-\bar{f}_y\bar{f}_s=\frac{1}{4}({\bar{f}}^2-f^2)\,,
\label{bilinear-SP2}
\end{eqnarray}
which are bilinear equations of the SP equation.

Next, we give the Casorati determinant ($N$-soliton)
solution of the SP equation.
It is known that the Casorati determinant solution of the
2DTL equation is of the form \cite{HIK,HirotaBook}:
\begin{equation}
\tau_n(Y,S)=\left|
\psi_{i}^{(n+j-1)}(Y,S)
\right|_{1\leq i,j\leq N}\,,
\end{equation}
where $\psi_i^{(n)}(Y,S)$ satisfies linear dispersive relations
\begin{equation}
  \frac{\partial \psi_i^{(n)}}{\partial Y}=\psi_{i}^{(n+1)},\quad
  \frac{\partial \psi_i^{(n)}}{\partial S}=\psi_{i}^{(n-1)}\,.
\end{equation}
For example, a particular choice of $\psi_i^{(n)}(Y,S)$
\begin{equation}
\psi_i^{(n)}(Y,S)
=c_{i,1}
p_i^ne^{p_iY+\frac{1}{p_i}S+\eta_{0i}}+
c_{i,2}q_i^ne^{q_iY+\frac{1}{q_i}S+\eta'_{0i}}\,,
\end{equation}
with $c_{i,1}$ and $c_{i,2}$ being constants,
satisfies the linear dispersive relations 
and gives the $N$-soliton solutions.

Applying the 2-reduction $q_i=-p_i$
and the change of variables $y=2Y$ and $s=2S$,
we obtain the determinant solution of bilinear equations
(\ref{bilinear-SP1}) and (\ref{bilinear-SP2}):
\[
f(y,s)=\tau_{0}(y,s), \quad \bar{f}(y,s)=\tau_{1}(y,s)\,,
\]
\begin{equation}
\tau_n(y,s)=\left|
\psi_{i}^{(n+j-1)}(y,s)
\right|_{1\leq i,j\leq N}\,,
\end{equation}
where
\begin{equation}
\psi_i^{(n)}(y,s)=c_{i,1}p_i^ne^{\frac{1}{2}p_iy+\frac{1}{2p_i}s+\eta_{0i}}
+c_{i,2}(-p_i)^ne^{-\frac{1}{2}p_iy-\frac{1}{2p_i}s+\eta'_{0i}}\,.
\end{equation}

Since $u$ is real and
the dependent variable transformation $u$ includes the imaginary
number, we must consider the reality condition of $u$.
Let us introduce $\alpha$ and $\beta$ such that
$F^*=\alpha \bar{f}$ and $F=\beta f$,
where $F$ and $F^*$ are complex conjugate of each other.
Note that $F$ and $F^*$ also
satisfies the bilinear equations (\ref{bilinear-SP1}) and 
(\ref{bilinear-SP2}) because of
\begin{equation}
\fl
u=\frac{\partial}{\partial s}\left(2\img \, \ln \frac{F^*}{F}\right)
=\frac{\partial}{\partial s}\left(2\img \, \ln
\frac{\alpha \bar{f}}{\beta f}\right)
=\frac{\partial}{\partial s}\left(2\img \, \ln \frac{\bar{f}}{f}+2\img
\, \ln \frac{\alpha}{\beta}\right)
=\frac{\partial}{\partial s}\left(2\img \, \ln \frac{\bar{f}}{f}\right)\,.
\end{equation}
Thus a set of $F$ and $F^*$ gives solutions of the SP equation as well as
a set of $f$ and $\bar{f}$.  By choosing phase constants appropriately,
the functions $f$ and $\bar{f}$ can be made to be complex
conjugate of each other to keep the reality and regularity of $u$.
For example, the following choice
\begin{equation}
\psi_i^{(n)}=p_i^ne^{\frac{1}{2}p_iy+\frac{1}{2p_i}s+\eta_{0i}-\img \pi/4}
+(-p_i)^ne^{-\frac{1}{2}p_iy-\frac{1}{2p_i}s+\eta'_{0i}+\img \pi/4}\,,
\end{equation}
guarantees the reality and regularity of the solution.

Summarizing the above results, the determinant ($N$-soliton)
solution of the SP equation is given by
\begin{equation}
u(y,s)
=\frac{\partial}{\partial s}\left(2\img \, \ln
\frac{\bar{f}(y,s)}{f(y,s)}\right)\,,
\end{equation}
\[
x=y-2\img (\ln \bar{f}f)_t, \quad t=s\,,
\]
\[
f(y,s)=\tau_{0}(y,s)\,, \quad
\bar{f}(y,s)=\tau_{1}(y,s) \,,
\]
\[
\tau_n(y,s)=\left|
\psi_{i}^{(n+j-1)}(y,s)
\right|_{1\leq i,j\leq N}\,\,,
\]
where
\[
\psi_i^{(n)}=p_i^ne^{\frac{1}{2}p_iy+\frac{1}{2p_i}s+\eta_{0i}-\img \pi/4}
+(-p_i)^ne^{-\frac{1}{2}p_iy-\frac{1}{2p_i}s+\eta'_{0i}+\img \pi/4}\,.
\]

\section{An integrable semi-discretization of the short pulse equation and 
numerical computations}

Based on the above fact, we construct the integrable
spatial-discretization of the SP equation. Consider the following
Casorati determinant:
\begin{equation}
\tau_n(k,S)=\left|
\psi_{i}^{(n+j-1)}(k,S)
\right|_{1\leq i,j\leq N}\,,
\end{equation}
where $\psi_i^{(n)}$ satisfies the dispersion relations
\begin{eqnarray}
&&\Delta_k\psi_i^{(n)} = \psi_i^{(n+1)}\,,
\label{k-dispersion-d}\\
&&
\partial_S\psi_i^{(n)} = -\psi_i^{(n-1)}\,.
\label{s-linear-d}
\end{eqnarray}
Here $\Delta_k$ is the backward difference operator with the spacing
constant $a$
\[
\Delta_k f(k)=\frac{f(k)-f(k-1)}{a} \,.
\]
Particularly, one can choose
\begin{equation}
\psi_i^{(n)}(k,S)=
c_{i,1}p_i^n(1-ap_i)^{-k}e^{\frac{1}{p_i}S+\xi_{i0}} +
c_{i,2}q_i^n(1-aq_i)^{-k}e^{\frac{1}{q_i}S+\eta_{i0}}\,,
\end{equation}
which automatically satisfies the dispersion relations (\ref{k-dispersion-d})
and (\ref{s-linear-d}). The
above Casorati determinant satisfies the bilinear form of the
semi-discrete 2DTL equation (the B\"acklund transformation of the
bilinear equation of 2DTL equation) \cite{OKMS,HirotaBook}
\begin{equation}
\left(\frac{1}{a}D_S
-1\right) \tau_n(k+1)\cdot \tau_n(k)
 +
\tau_{n+1}(k+1)\tau_{n-1}(k)= 0\,.
\label{2DTL}
\end{equation}
Applying 2-reduction
$$
q_i=-p_i\,,
$$
and letting
$$
f_k=\tau_0(k)\,,\quad \bar{f}_k=\tau_{1}(k)
=\left(\prod_{i=1}^Np_i^2\right)\tau_{-1}(k)\,,
$$
we obtain 
\begin{eqnarray}
&&  \frac{1}{a}D_S f_{k+1}\cdot 
f_k
-f_{k+1}f_k
+
\bar{f}_{k+1}\bar{f}_k= 0\,,\\
&& \frac{1}{a}D_S \bar{f}_{k+1}\cdot
\bar{f}_k
-\bar{f}_{k+1}\bar{f}_k
+
{f}_{k+1}{f}_k= 0\,,
\end{eqnarray}
where the gauge transformation 
$\tau_n \to \left(\prod_{i=1}^Np_i\right)^{n}\tau_n$ is used. 
Note that $f$ and $\bar{f}$ can be made complex conjugate of each other 
by choosing the phase constants properly. 
Under the change of independent variable
$s=2S$,
Eq.(\ref{2DTL}) implies the following two bilinear equations
\begin{eqnarray}
&&\frac{2}{a}D_s f_{k+1}\cdot
f_k
-f_{k+1}f_k
+
\bar{f}_{k+1}\bar{f}_k= 0\,,\\
&&\frac{2}{a}D_s \bar{f}_{k+1}\cdot
\bar{f}_k
-\bar{f}_{k+1}\bar{f}_k
+
{f}_{k+1}{f}_{k}= 0\,,
\end{eqnarray}
which can be readily shown to be equivalent to
\begin{eqnarray}
&&-\left(\frac{2}{a}
\left(\ln \frac{f_{k+1}}{f_k}\right)_s
-1\right)
= \frac{\bar{f}_{k+1}\bar{f}_k}{f_{k+1}f_k}\,,\\
&&-\left(\frac{2}{a}
\left(\ln \frac{\bar{f}_{k+1}}{\bar{f}_k}\right)_s
-1\right)
= \frac{{f}_{k+1}{f}_k}{\bar{f}_{k+1}\bar{f}_k}\,.
\end{eqnarray}
Subtracting the above two equations, one obtains
\begin{equation}
\frac{2}{a}\left(
\left(\ln \frac{\bar{f}_{k+1}}{\bar{f}_k}\right)_s
-\left(\ln \frac{f_{k+1}}{f_k}\right)_s
\right)
=\frac{\bar{f}_{k+1}\bar{f}_k}{f_{k+1}f_k} 
-\frac{{f}_{k+1}f_k}{\bar{f}_{k+1}\bar{f}_k}\,.
\end{equation}
Introducing the dependent variable transformation
$\phi_k(s)=2{\rm i}\ln\left(\frac{\bar{f}_k(s)}{f_k(s)}\right)$, 
one arrives at
\begin{equation}\label{semi-SG}
\frac{\phi_{k+1,s}-\phi_{k,s}}{2a}=\sin 
\left(\frac{\phi_{k+1}+\phi_k}{2}\right)\,,
\end{equation}
which is nothing but an integrable semi-discretization of the
sine-Gordon equation.
Note that this is also known as the B\"acklund transformation 
of the sine-Gordon equation \cite{Backlund,Hirota-SG}. 

It is obvious that, from the semi-discrete sine-Gordon equation
(\ref{semi-SG}), the equation
\begin{equation}
\left(\cos \left(\frac{\phi_{k+1}+\phi_k}{2}\right)\right)_s
=-\frac{\phi_{k+1,s}^2-\phi_{k,s}^2}{4a}\,,
\end{equation}
is implied.
By introducing the dependent variable transformation
\begin{equation}\label{u-phi-relation}
u_k=\frac{d \phi_{k}}{d s}
= 2{\rm i}\ln\left(\frac{\bar{f}_k(s)}{f_k(s)}\right)_s\,,
\quad 
\delta_k=\cos \left(\frac{\phi_{k+1}+\phi_k}{2}\right)\,,
\end{equation}
it then follows 
\begin{equation}
\frac{d \delta_{k}}{d s}= -\frac{u_{k+1}^2-u_k^2}{4}\,,
\label{sd-SP1}
\end{equation}
which is the first equation of a semi-discrete 
analogue of the SP equation

From the facts
\begin{equation}
\cos^2
\left(\frac{\phi_{k+1}+\phi_k}{2}\right)
+\sin^2 
\left(\frac{\phi_{k+1}+\phi_k}{2}\right)
=1\,,
\end{equation}
\begin{equation}\label{sin-equation}
\sin
\left(\frac{\phi_{k+1}+\phi_k}{2}\right)=\frac{u_{k+1}-u_k}{2a} \,,
\end{equation}
and 
\begin{equation}
\frac{1}{r_k}=\frac{\delta_k}{a}=\cos 
\left(\frac{\phi_{k+1}+\phi_k}{2}\right) \,,
\end{equation}
it follows
\[
\frac{\delta_k^2}{a^2}+\frac{(u_{k+1}-u_k)^2}{4a^2}=1\,, 
\]
i.e., 
\begin{equation}
\delta_k^2=a^2-\frac{(u_{k+1}-u_k)^2}{4}\,,
\end{equation}
which becomes another equation of a semi-discrete analogue of 
the SP equation. 

Summarizing the above results, 
we obtained an integrable semi-discrete analogue of 
the SP equation and its solutions
\begin{eqnarray}
&&(u_{k+1}-u_k)^2 =4(a^2-\delta_k^2)\,,\label{sd-SP1}\\
&&\frac{d \delta_{k}}{d s}= -\frac{u_{k+1}^2-u_k^2}{4}\,,
\label{sd-SP2}
\end{eqnarray}
where the $x$-coordinate of the $k$-th lattice point 
is given by $X_k=X_0+\sum_{l=0}^{k-1}\delta_l$. 
From the construction, 
the semi-discrete analogue of the SP equation has the following
Casorati determinant solution:
\begin{equation}
u_k(s)=\frac{d}{d s}
\left(2{\rm i}\ln\, \frac{\bar{f}_k}{f_k}\right)\,,\quad  
\delta_k=\frac{a}{2}
\left(\frac{\bar{f}_{k+1}\bar{f}_{k}}{f_{k+1}f_k}
+\frac{{f}_{k+1}{f}_{k}}{\bar{f}_{k+1}\bar{f}_k}
\right)\,, 
\end{equation}
$$X_k=X_0+\sum_{l=0}^{k-1}\delta_l\,,$$
$$f_k(s)=\tau_0(k,s)\,,\quad \bar{f}_k(s)=\tau_{1}(k,s)\,,$$
$$
\tau_n(k,s)=\left|
\psi_{i}^{(n+j-1)}(k,s)
\right|_{1\leq i,j\leq N}\,,
$$
where $\psi_i^{(n)}(k,s)$ satisfies
\[
\fl 
\psi_i^{(n)}(k,s)=p_i^n(1-ap_i)^{-k}e^{\frac{1}{2p_i}s+\xi_{i0}-\img \pi/4}
+(-p_i)^n(1+ap_i)^{-k}e^{-\frac{1}{2p_i}s+\eta_{i0}+\img \pi/4}\,,
\]
and the phase constants $\pm \img \pi/4$ play a role of 
keeping the reality and regularity. 

Note that $a^2$ must be always greater than or equal to $\delta_k^2$ 
because $(u_{k+1}-u_k)^2\geq 0$. 
This can be easily verified by
\begin{equation}
|\delta_k|=
\left|a\,\cos\left(\frac{\phi_{k+1}+\phi_k}{2}\right)\right|\leq |a|\,. 
\end{equation}
The mesh size of self-adaptive mesh $|\delta_k|$ is 
always chosen as less than $|a|$.

We can rewrite the semi-discrete SP equation in an 
alternative form which converges to 
the SP equation in the continuous limit $\delta_k\to 0$. 
Multiplying Eq.(\ref{sd-SP2}) by 2$\delta_k$, we have 
\begin{equation}
\frac{d \delta_{k}^2}{d s}
= -\delta_k\frac{u_{k+1}^2-u_k^2}{2}\,.
\end{equation}
Eliminating $\delta_k^2$ using Eq.(\ref{sd-SP1}), this leads to
\begin{equation}
\frac{d (u_{k+1}-u_k)}{d s}
=\delta_k (u_{k+1}+u_k)\,.\label{semi-SP-mod}
\end{equation}
Since
\begin{equation}
\frac{d}{d s}\left(
\frac{u_{k+1}-u_k}{\delta_k }
\right)=\frac{1}{\delta_k} {\frac{d (u_{k+1}-u_k)}{d s}}
-\frac{u_{k+1}-u_k}{\delta_k^2}\frac{d \delta_k}{d s}\,,
\end{equation}
it follows that
\begin{equation} \label{sd-SP2-2}
\frac{d}{d s}\left(
\frac{u_{k+1}-u_k}{\delta_k }
\right)=u_{k+1}+u_k+\frac{u_{k+1}+u_k}{4}
\left(\frac{u_{k+1}-u_k}{\delta_k}\right)^2\,,
\end{equation}
by using Eqs.(\ref{semi-SP-mod}) and (\ref{sd-SP2}). 
Equation (\ref{sd-SP2-2})
gives another form of the semi-discrete SP equation.
In the continuous limit $a\to0$ ($\delta_k\to0$), we have
$$
\frac{u_{k+1}-u_k}{\delta_k}\to\frac{d u}{d x}\,,
\qquad\frac{u_{k+1}+u_k}{2}\to u\,,
$$
$$
\frac{\partial x}{\partial s}
=\frac{\partial x_0}{\partial s}
+\sum_{j=0}^{k-1}\frac{d\delta_j}{d s}
= -\frac 14\sum_{j=0}^{k-1}(u^2_{j+1}-u^2_j)
\to -\frac 14 u^2\,,
$$
$$
\partial_s=\partial_t+\frac{\partial x}{\partial s}\partial_x
\to\partial_t  -\frac 14 u^2\partial_x\,,
$$
Consequently, Eq.(\ref{sd-SP2-2}) converges to
$$
\left(\partial_t  -\frac 14 u^2\partial_x\right) u_x 
= 2u+ \frac{1}{2}uu^2_x.
$$
By the scaling transformation $2x\to x$, one arrives
\[
u_{xt}=u+u(u_x)^2+\frac{1}{2}u^2u_{xx}\,,
\]
which turns out to be the SP equation
\[
 u_{xt}=u+\frac{1}{6}(u^3)_{xx}\,.
\]

\begin{figure}[htbp]
\centerline{
\includegraphics[scale=0.35]{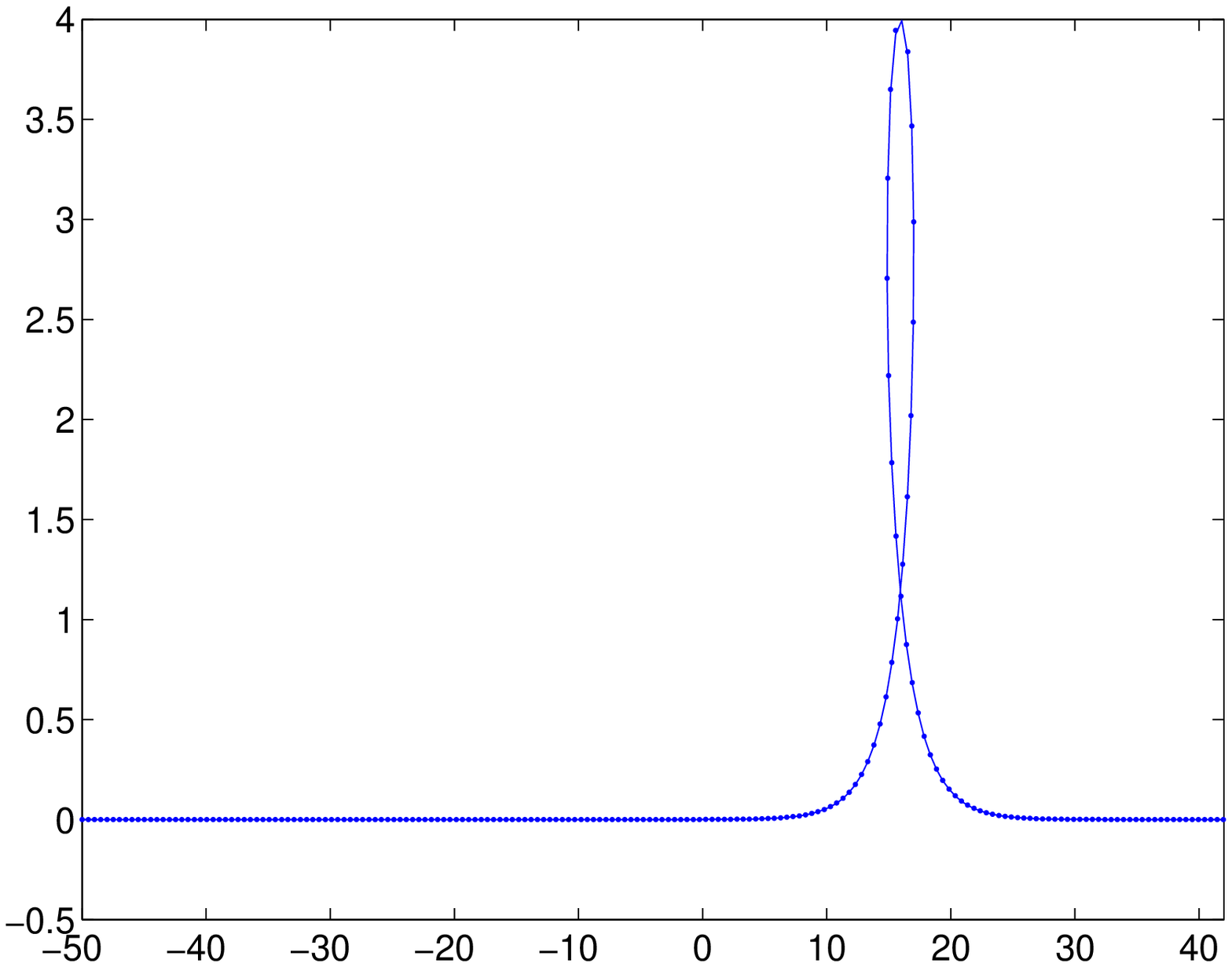}\quad
\includegraphics[scale=0.35]{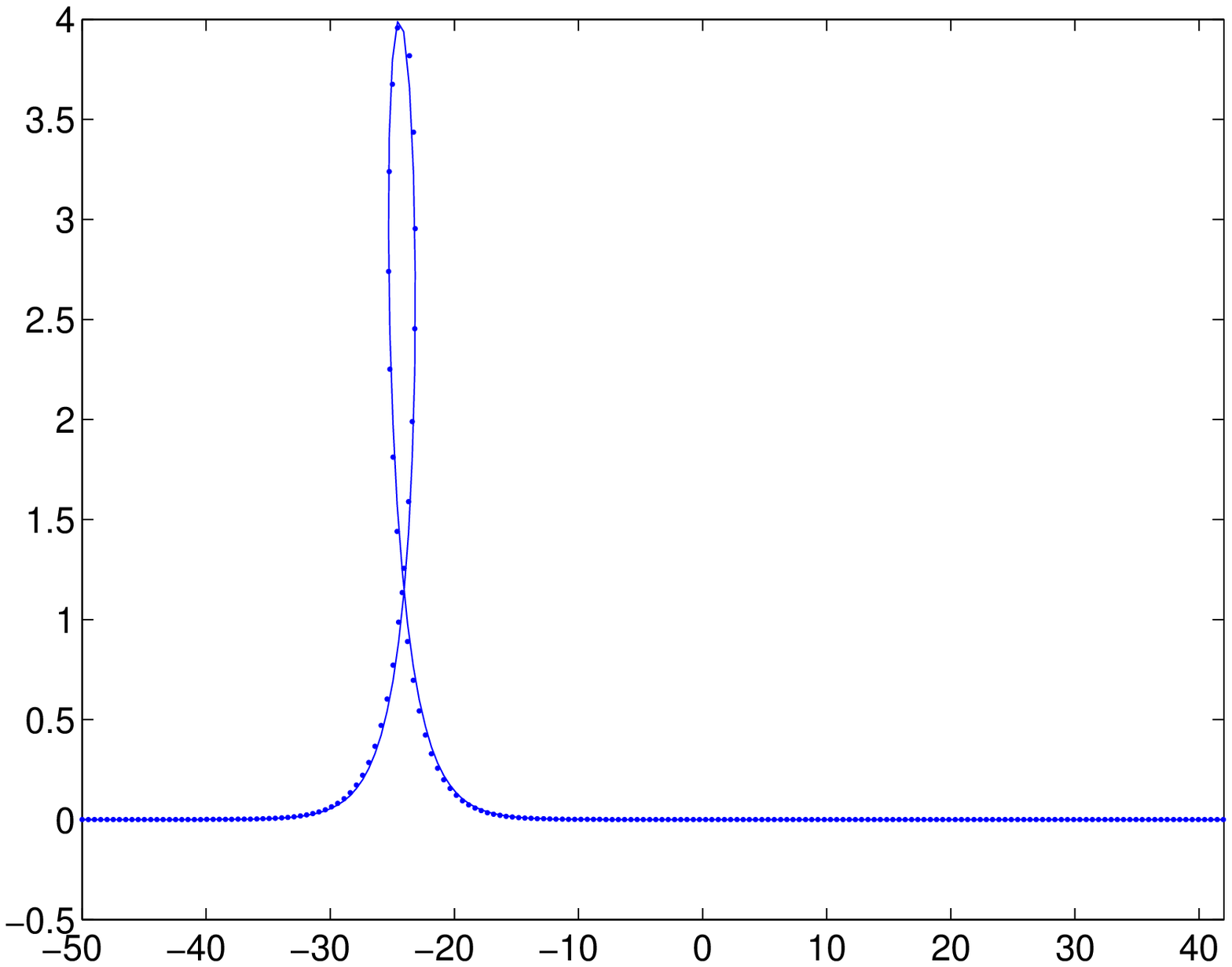}}
\kern-0.3\textwidth \hbox to
\textwidth{\hss(a)\kern9em\hss(b)\kern15em}
\kern+0.3\textwidth
\caption{Numerical solutions for
one-loop soliton solution with 
(a) $t=0.0$; (b) $t=10.0$. The parameters of the initial condition are 
$p_1=0.5$} \label{f:1loop}
\end{figure}

\begin{figure}[htbp]
\centerline{
\includegraphics[scale=0.35]{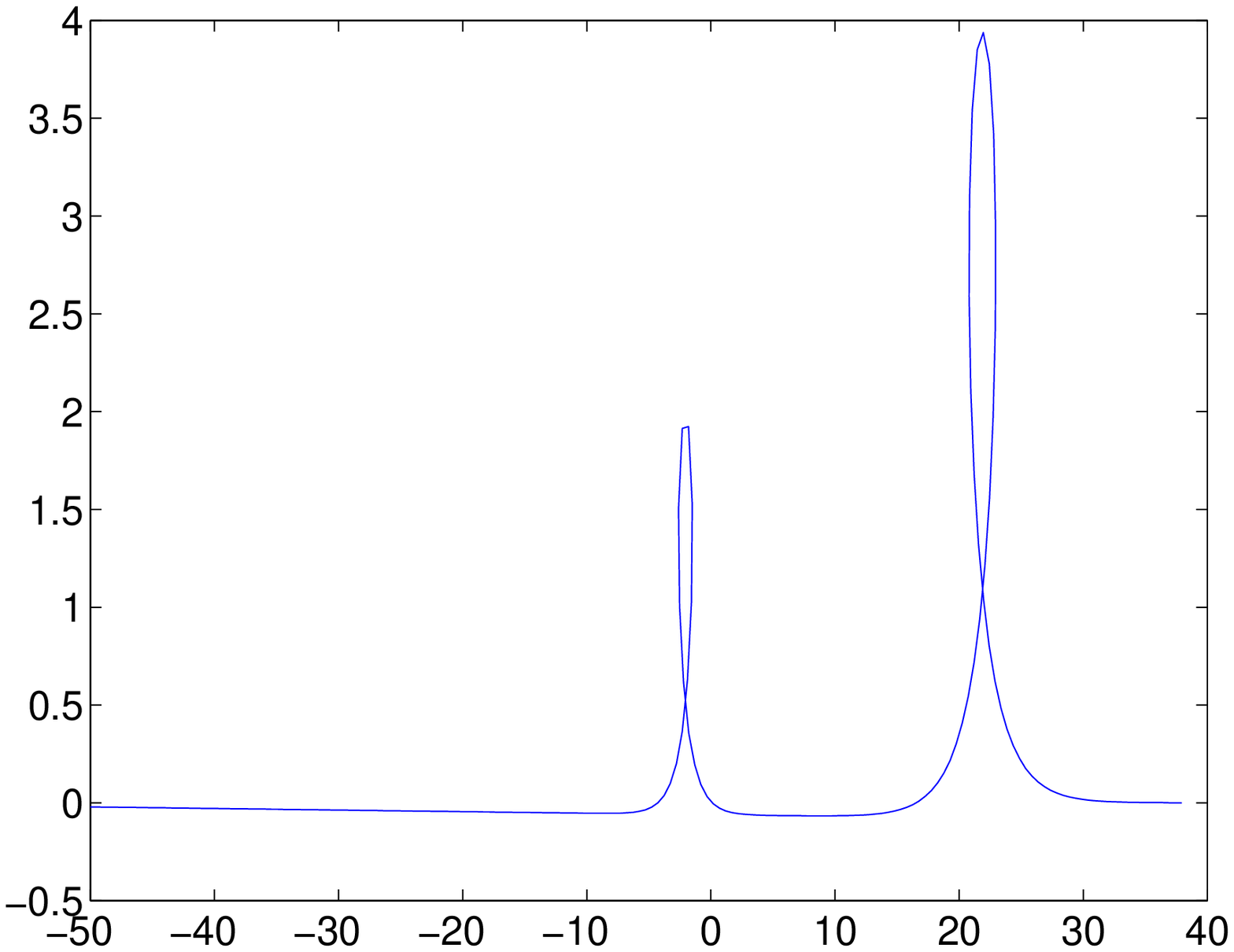}\quad
\includegraphics[scale=0.35]{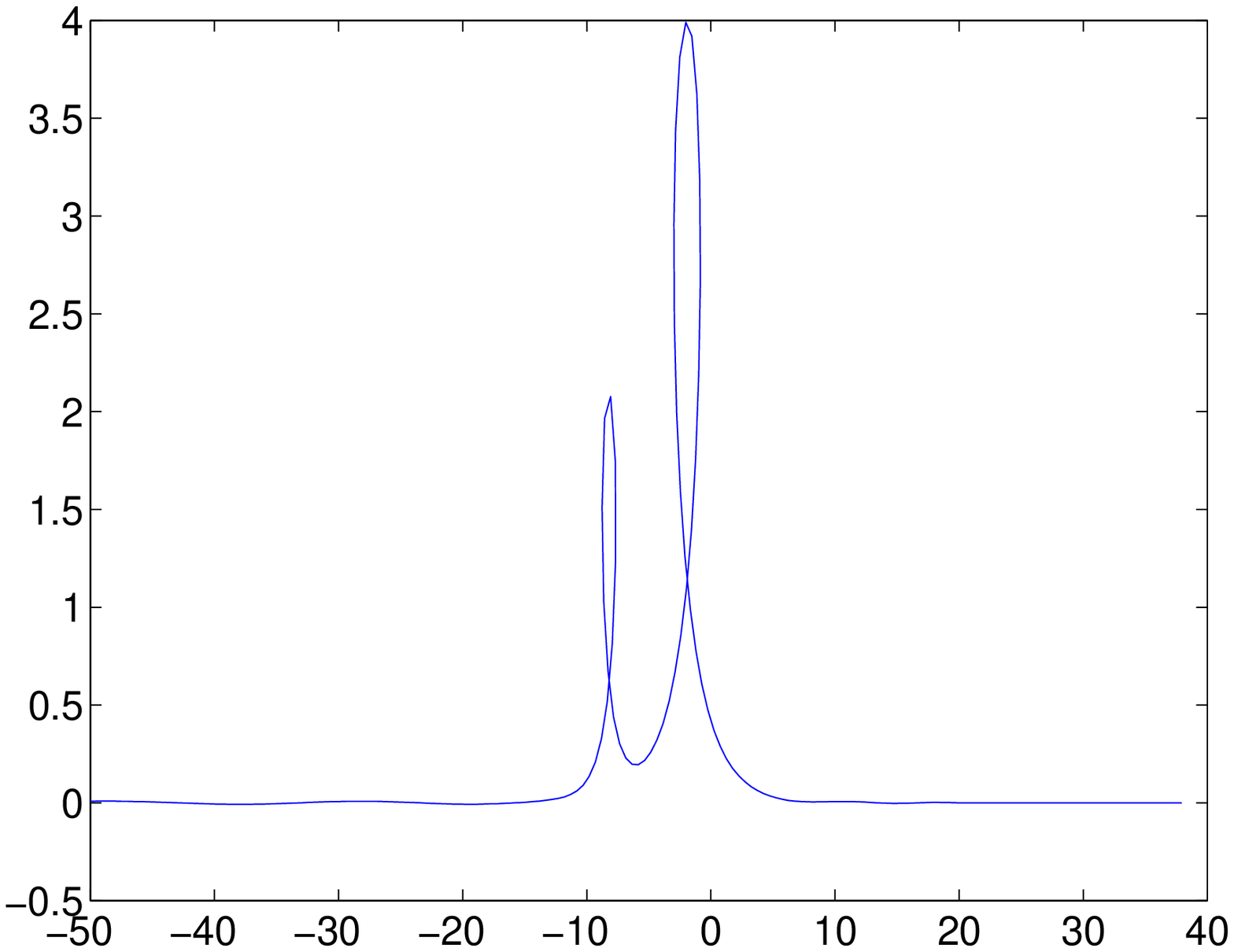}}
\kern-0.3\textwidth \hbox to
\textwidth{\hss(a)\kern9em\hss(b)\kern15em}
\kern+0.3\textwidth\centerline{
\includegraphics[scale=0.35]{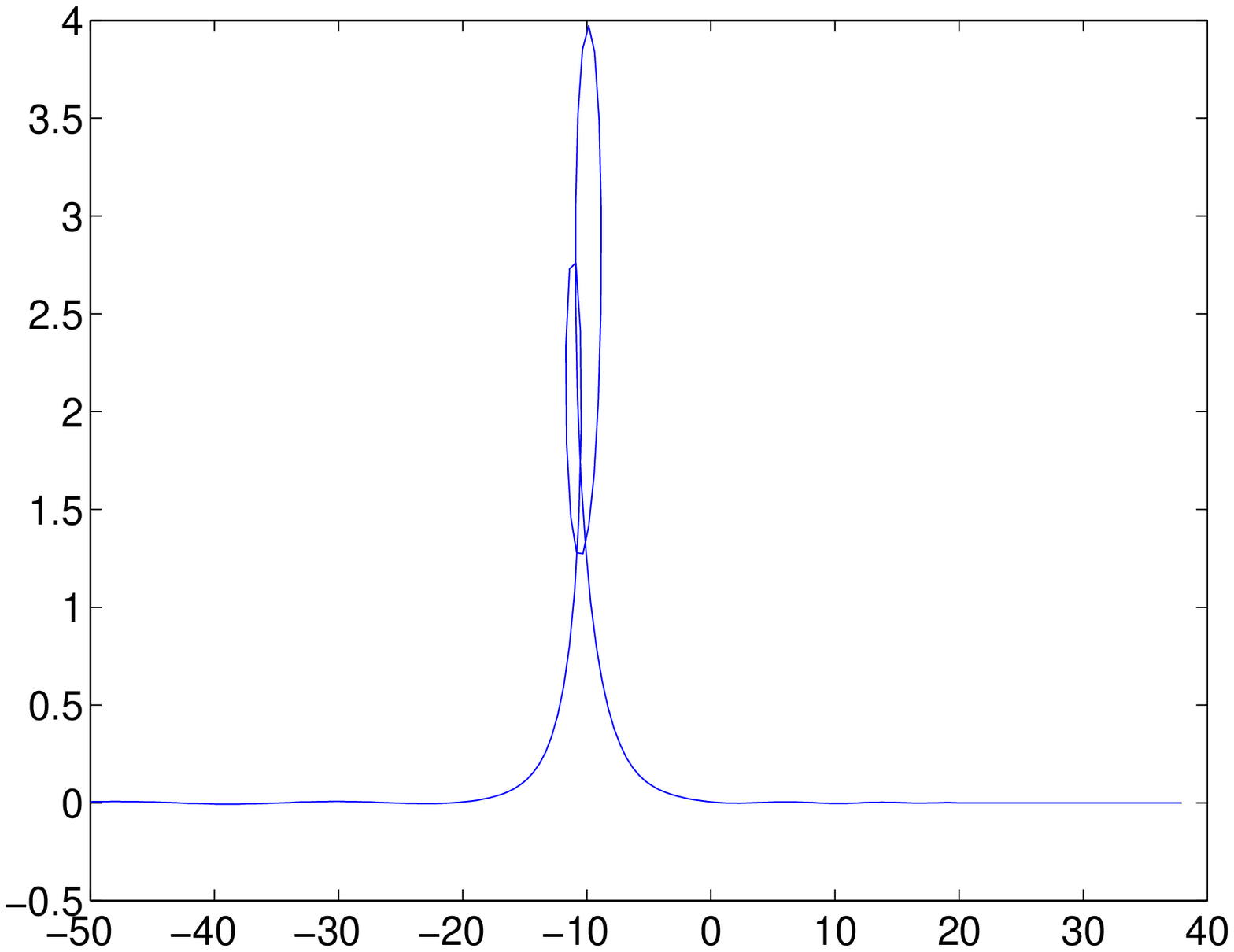}\quad
\includegraphics[scale=0.35]{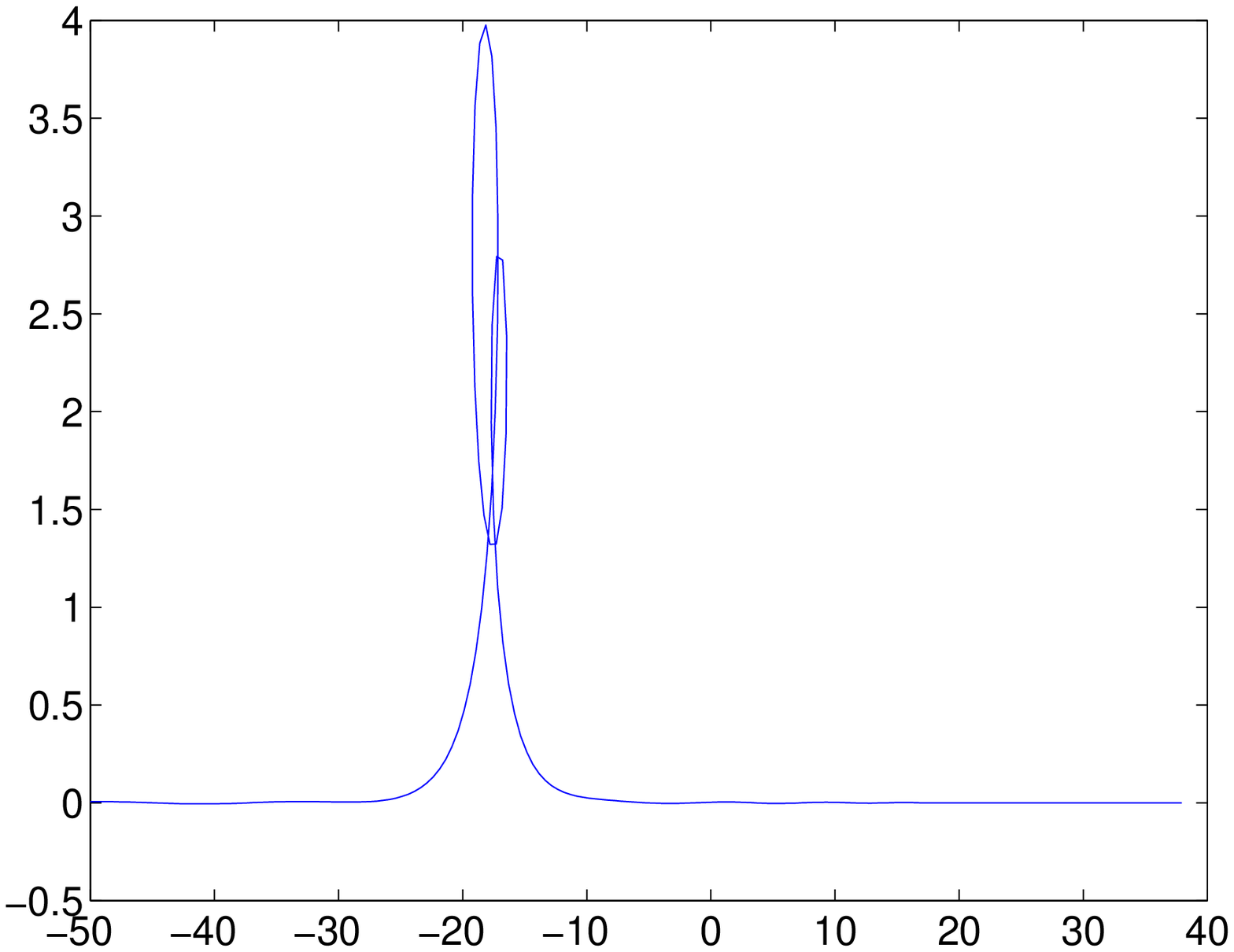}}
\kern-0.3\textwidth \hbox to
\textwidth{\hss(c)\kern9em\hss(d)\kern15em} 
\kern+0.3\textwidth
\centerline{
\includegraphics[scale=0.35]{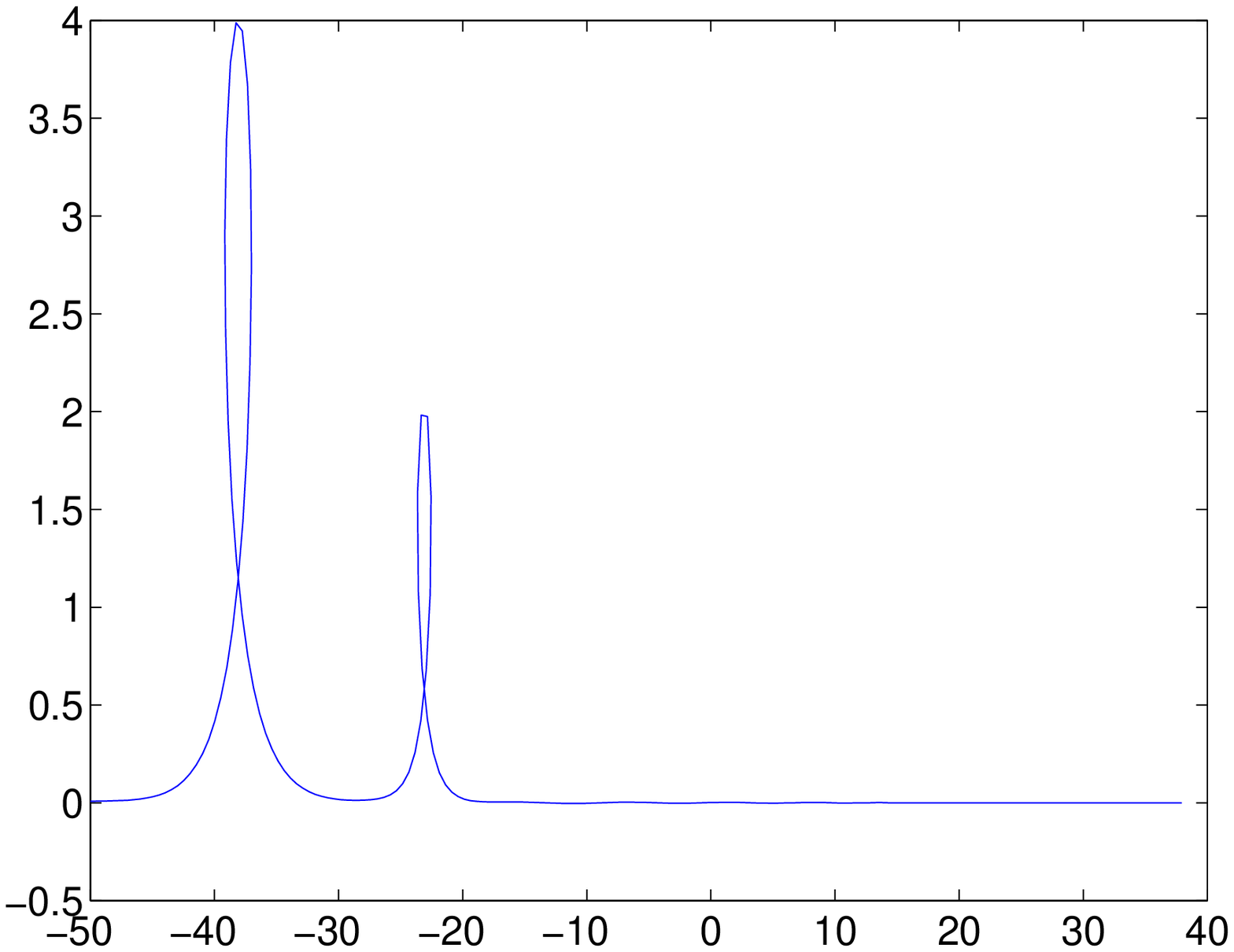}}
\kern-0.3\textwidth \hbox to \textwidth{\hss(e)\kern13em}
\kern+0.3\textwidth
\caption{Numerical solutions for
the collision of two-loop soliton solution with 
(a) $t=0.0$; (b) $t=6.0$; (c)
 $t=8.0$; (d) $t=10.0$; (e) $t=15.0$. 
The parameters of the initial condition 
are $p_1=0.5$, $p_2=1.0$.} \label{f:2loop}
\end{figure}

In a similar way employed in \cite{dCH,dCH2}, the semi-discrete
analogue of the SP equation can be used as a novel numerical scheme,
i.e., the so-called self-adaptive moving mesh method, to perform
numerical computations for the SP equation. However, the first
equation (\ref{sd-SP1}) 
has ambiguity for determining the sign even if the
non-uniform mesh $\delta_k$ is solved from the second equation 
(\ref{sd-SP2}). To
avoid this difficulty, we introduce an intermediate variable
$\bar{\phi}_k = (\phi_{k+1}+\phi_k)/2$, and employ the following
scheme,
\begin{equation} \label{sam}
    \left\{\begin{array}{l}\displaystyle
    (u_{k+1}-u_k) =2a \sin (\bar{\phi}_k),
    \\[5pt]\displaystyle
    \frac{d \bar{\phi}_k}{d s}= \frac{u_{k+1}+u_k}{2}.
    \end{array} \right.
\end{equation}
which can be derived from Eqs.(\ref{sin-equation}) 
and (\ref{u-phi-relation}). 
Equations (\ref{sam}) are equivalent to the integrable semi-discrete
analogue of the SP equation, and the relation between the
non-uniform mesh $\delta_k$ and $\bar{\phi}_k$ is $ \delta_k = a
\cos(\bar{\phi}_k)$.
Figures \ref{f:1loop} and \ref{f:2loop}
are numerical results for one-loop and two-loop soliton solutions,
respectively.
The time stepsize is $\Delta t=0.01$ and the number of grid points is $N=200$.
The detailed numerical results by using the integrable 
semi-discrete SP equation
will be reported somewhere else.

\section{Full-discretizations of the short pulse equation}

To construct a full-discrete analogue of the SP equation, 
we introduce one more discrete variable $l$ which corresponds to 
the discrete time variable. 

It is known that the $\tau$-function
\begin{equation}
\tau_n(k,l)=\left|\psi_i^{(n+j-1)}(k,l)\right|_{1\le i,j\le N}\,,
\end{equation}
with
\[
\fl 
\psi_i^{(n)}(k,l)
=c_{i,1}p_i^n(1-ap_i)^{-k}\left(1-b\frac{1}{p_i}\right)^{-l}
e^{\frac{1}{2p_i}s+\xi_{i0}}
+c_{i,2}q_i^n(1-aq_i)^{-k}\left(1-b\frac{1}{q_i}\right)^{-l}
e^{\frac{1}{2q_i}s+\eta_{i0}}\,,
\]
satisfies bilinear equations \cite{OKMS}
\begin{equation} \label{Backlund1}
\left(\frac{2}{a}D_s-1\right)\tau_n(k+1,l)\cdot\tau_n(k,l)
+\tau_{n+1}(k+1,l)\tau_{n-1}(k,l)=0\,,
\end{equation}
and 
\begin{equation} \label{Backlund2}
(2bD_s-1)\tau_{n}(k,l+1)\cdot\tau_{n+1}(k,l)
+\tau_{n}(k,l)\tau_{n+1}(k,l+1)=0\,.
\end{equation}
Applying the 2-reduction 
$\tau_{n-1}=\left(\prod_{i=1}^Np_i^2\right)^{-1} \tau_{n+1}$, 
i.e., adding constraints $q_i=-p_i$ to the $N$-soliton solution,
we obtain
\begin{equation}\label{Backlund1-2}
\left(\frac{2}{a}D_s-1\right)\tau_n(k+1,l)\cdot\tau_n(k,l)
+\tau_{n+1}(k+1,l)\tau_{n+1}(k,l)=0\,,
\end{equation}
and 
\begin{equation}\label{Backlund2-2}
(2bD_s-1)\tau_{n}(k,l+1)\cdot\tau_{n+1}(k,l)
+\tau_{n}(k,l)\tau_{n+1}(k,l+1)=0\,,
\end{equation}
where the gauge transformation $\tau_n \to 
\left(\prod_{i=1}^Np_i\right)^{n}\tau_n$ is used. 
Letting
$$
f_{k,l}=\tau_0(k,l)\,,\quad \bar{f}_{k,l}=\tau_{1}(k,l)\,,
$$
the bilinear equations (\ref{Backlund1-2}) and
(\ref{Backlund2-2}) imply the following four equations
\begin{eqnarray}
&&\left(\frac{2}{a}D_s-1\right)f_{k+1,l}\cdot f_{k,l}
+\bar{f}_{k+1,l}\bar{f}_{k,l}=0\,,\\
&&\left(\frac{2}{a}D_s-1\right)\bar{f}_{k+1,l}\cdot \bar{f}_{k,l}
+{f}_{k+1,l}{f}_{k,l}=0\,,\\
&&(2bD_s-1)f_{k,l+1}\cdot \bar{f}_{k,l}
+f_{k,l}\bar{f}_{k,l+1}=0\,,\\
&&(2bD_s-1)\bar{f}_{k,l+1}\cdot f_{k,l}
+\bar{f}_{k,l}f_{k,l+1}=0\,,
\end{eqnarray}
which are actually equivalent to
\begin{eqnarray}
&&\frac{2}{a}\left(\ln\frac{f_{k+1,l}}{f_{k,l}}\right)_s-1
+\frac{\bar{f}_{k+1,l}\bar{f}_{k,l}}{f_{k+1,l}f_{k,l}}=0\,,\\
&&\frac{2}{a}\left(\ln\frac{\bar{f}_{k+1,l}}{\bar{f}_{k,l}}\right)_s-1
+\frac{f_{k+1,l}f_{k,l}}{\bar{f}_{k+1,l}\bar{f}_{k,l}}=0\,,\\
&&2b\left(\ln\frac{f_{k,l+1}}{\bar{f}_{k,l}}\right)_s-1
+\frac{f_{k,l}\bar{f}_{k,l+1}}{f_{k,l+1}\bar{f}_{k,l}}=0\,,\\
&&2b\left(\ln\frac{\bar{f}_{k,l+1}}{f_{k,l}}\right)_s-1
+\frac{\bar{f}_{k,l}f_{k,l+1}}{\bar{f}_{k,l+1}f_{k,l}}=0\,.
\end{eqnarray}
Note that $f$ and $\bar{f}$ can be made complex conjugate of each other 
by choosing the phase constants properly. 
By introducing
\begin{equation}
u_{k,l}=\left(2{\rm i}\ln\frac{\bar{f}_{k,l}}{f_{k,l}}\right)_s\,,
\end{equation}
and 
\begin{equation}
X_{k,l}=ka-(\ln\bar{f}_{k,l}f_{k,l})_s\,,
\end{equation}
where $X_{k,l}$ is the $x$-coordinate of the $k$-th lattice point at 
time $l$,  
we find the following relations
\begin{eqnarray}
&&u_{k+1,l}-u_{k,l}
={\rm i}a\left(\frac{\bar{f}_{k+1,l}\bar{f}_{k,l}}{f_{k+1,l}f_{k,l}}
-\frac{f_{k+1,l}f_{k,l}}{\bar{f}_{k+1,l}\bar{f}_{k,l}}\right)\,,
\label{u-k}\\
&&u_{k,l+1}+u_{k,l}
=\frac{{\rm i}}{b}\left(\frac{f_{k,l}\bar{f}_{k,l+1}}{f_{k,l+1}\bar{f}_{k,l}}
-\frac{\bar{f}_{k,l}f_{k,l+1}}{\bar{f}_{k,l+1}f_{k,l}}\right)\,,
\label{u-l}\\
&&X_{k+1,l}-X_{k,l}
=\frac{a}{2}\left(\frac{\bar{f}_{k+1,l}\bar{f}_{k,l}}{f_{k+1,l}f_{k,l}}
+\frac{f_{k+1,l}f_{k,l}}{\bar{f}_{k+1,l}\bar{f}_{k,l}}\right)\,,
\label{X-k}\\
&&X_{k,l+1}-X_{k,l}
=-\frac{1}{b}
+\frac{1}{2b}\left(\frac{f_{k,l}\bar{f}_{k,l+1}}{f_{k,l+1}\bar{f}_{k,l}}
+\frac{\bar{f}_{k,l}f_{k,l+1}}{\bar{f}_{k,l+1}f_{k,l}}\right)\,.\label{X-l}
\end{eqnarray}

It is straightforward to derive
\begin{equation}
(u_{k+1,l}-u_{k,l})^2=4(a^2-\delta_{k,l}^2)\,,\label{u-k-fulldiscrete}
\end{equation}
from Eqs.(\ref{u-k}) and (\ref{X-k})
and
\begin{equation}
(u_{k,l+1}+u_{k,l})^2=4\left(\frac{1}{b^2}-
\left(X_{k,l+1}-X_{k,l}+\frac{1}{b}\right)^2\right)\,. 
\label{u-l-fulldiscrete}
\end{equation}
from Eqs.(\ref{u-l}) and (\ref{X-l}),
where $\delta_{k,l}=X_{k+1,l}-X_{k,l}$.
Equations (\ref{u-k-fulldiscrete}) and (\ref{u-l-fulldiscrete}) 
give a full-discrete analogue of the SP equation. 

Let us consider another full-discrete analogue of 
the SP equation. 
From Eqs.(\ref{u-k})-Eqs.(\ref{X-l}), we obtain
\begin{eqnarray}
&&\frac{\bar{f}_{k+1,l}\bar{f}_{k,l}}{f_{k+1,l}f_{k,l}}
=\frac{1}{a}\left(X_{k+1,l}-X_{k,l}-{\rm i}\frac{u_{k+1,l}-u_{k,l}}{2}\right)\,,
\label{X-u-1}\\
&&\frac{f_{k+1,l}f_{k,l}}{\bar{f}_{k+1,l}\bar{f}_{k,l}}
=\frac{1}{a}\left(X_{k+1,l}-X_{k,l}+{\rm i}
\frac{u_{k+1,l}-u_{k,l}}{2}\right)\,,\\
&&\frac{f_{k,l}\bar{f}_{k,l+1}}{f_{k,l+1}\bar{f}_{k,l}}
=b\left(X_{k,l+1}-X_{k,l}+\frac{1}{b}-{\rm i}
\frac{u_{k,l+1}+u_{k,l}}{2}\right)\,,\\
&&\frac{\bar{f}_{k,l}f_{k,l+1}}{\bar{f}_{k,l+1}f_{k,l}}
=b\left(X_{k,l+1}-X_{k,l}+\frac{1}{b}+{\rm i}
\frac{u_{k,l+1}+u_{k,l}}{2}\right)\,.
\label{X-u-4}
\end{eqnarray}
From the relations (\ref{X-u-1})-(\ref{X-u-4}), we have 
\begin{eqnarray}
&&\frac{X_{k+1,l+1}-X_{k,l+1}-{\rm i}\frac{u_{k+1,l+1}-u_{k,l+1}}{2}}
{X_{k+1,l}-X_{k,l}-{\rm i}\frac{u_{k+1,l}-u_{k,l}}{2}}\nonumber\\
&&\qquad 
=\frac{X_{k+1,l+1}-X_{k+1,l}+\frac{1}{b}-{\rm i}\frac{u_{k+1,l+1}+u_{k+1,l}}{2}}
{X_{k,l+1}-X_{k,l}+\frac{1}{b}+{\rm i}\frac{u_{k,l+1}+u_{k,l}}{2}}\,.
\end{eqnarray}
Equating the real part and imaginary part respectively, we have
\begin{eqnarray}
\fl &&(X_{k+1,l+1}-X_{k,l+1})\left(X_{k,l+1}-X_{k,l}+\frac{1}{b}\right)
+\frac{u_{k+1,l+1}-u_{k,l+1}}{2}\frac{u_{k,l+1}+u_{k,l}}{2}\nonumber\\
\fl &&\quad =\left(X_{k+1,l+1}-X_{k+1,l}+\frac{1}{b}\right)(X_{k+1,l}-X_{k,l})
-\frac{u_{k+1,l+1}+u_{k+1,l}}{2}\frac{u_{k+1,l}-u_{k,l}}{2}\,,\label{fully-SP1}\\
\fl &&\left(X_{k,l+1}-X_{k,l}+\frac{1}{b}\right)(u_{k+1,l+1}-u_{k,l+1})
-(X_{k+1,l+1}-X_{k,l+1})(u_{k,l+1}+u_{k,l})\nonumber\\
\fl &&\quad =\left(X_{k+1,l+1}-X_{k+1,l}+\frac{1}{b}\right)(u_{k+1,l}-u_{k,l})
+(X_{k+1,l}-X_{k,l})(u_{k+1,l+1}+u_{k+1,l})\,,\label{fully-SP2}
\end{eqnarray}
which can be rearranged into the
following simpler form:
\begin{eqnarray}
&&(X_{k+1,l+1}-X_{k+1,l}-X_{k,l+1}+X_{k,l})
\left(\frac{1}{b}-X_{k+1,l}+X_{k,l+1}\right)
\nonumber\\
&&\qquad =-\frac{u_{k+1,l+1}+u_{k+1,l}-u_{k,l+1}-u_{k,l}}{2}
\frac{u_{k+1,l}+u_{k,l+1}}{2}\,,
\label{eqX}\\
&&(u_{k+1,l+1}-u_{k+1,l}-u_{k,l+1}+u_{k,l})
\left(\frac{2}{b}+X_{k+1,l+1}-X_{k+1,l}+X_{k,l+1}-X_{k,l}\right)
\nonumber\\
&&\, =(X_{k+1,l+1}+X_{k+1,l}-X_{k,l+1}-X_{k,l})
(u_{k+1,l+1}+u_{k+1,l}+u_{k,l+1}+u_{k,l})\,.
\label{equ}
\end{eqnarray}
Equations (\ref{eqX}) and (\ref{equ}) constitute 
another form of integrable full-discretization of the SP equation.
Taking the continuous limit $b\to 0$ in time, 
we obtain 
\begin{equation}
(X_{k+1}-X_k)_s=-\frac{1}{4}(u_{k+1}-u_{k})(u_{k+1}+u_k)\,,
\end{equation}
and 
\begin{equation}
(u_{k+1}-u_k)_s=(X_{k+1}-X_k)(u_{k+1}+u_k)\,,
\end{equation}
which are nothing but the semi-discrete analogue of the SP equation
(\ref{sd-SP1}) and (\ref{sd-SP2}). 
Here we used $\frac{F_{l+1}-F_l}{2b}\to\partial_sF$ as $b\to 0$. 

From the construction of the full-discrete analogue of the SP equation, 
the determinant solution of the full-discrete SP equation is 
\begin{eqnarray}
&&u_{k,l}
={\rm i} \left(\frac{\bar{g}_{k,l}}{\bar{f}_{k,l}}-
\frac{{g}_{k,l}}{{f}_{k,l}}
\right)
=\frac{\partial}{\partial s}
\left(2{\rm i}\ln\, \frac{\bar{f}_{k,l}}{f_{k,l}}\right)
\,, \\
&& X_{k,l}
=ka-\frac{1}{2}
\left(\frac{\bar{g}_{k,l}}{\bar{f}_{k,l}}+\frac{{g}_{k,l}}{{f}_{k,l}}
\right)
=ka-\frac{\partial}{\partial s}(\ln\bar{f}_{k,l}f_{k,l})
\,,
\end{eqnarray}
$$f_{k,l}=\tau_0(k,l)\,,\quad 
\bar{f}_{k,l}=\tau_{1}(k,l)\,,$$
$$g_{k,l}=\rho_0(k,l)\,,\quad 
\bar{g}_{k,l}=\rho_{1}(k,l)\,,$$
$$
\tau_n(k,l)=\left|
\begin{array}{cccc}
\psi_1^{(n)}(k,l) & \psi_1^{(n+1)}(k,l) & \cdots & 
\psi_1^{(n+N-1)}(k,l)\\
\psi_2^{(n)}(k,l) & \psi_2^{(n+1)}(k,l) & \cdots &  
\psi_2^{(n+N-1)}(k,l)\\
\vdots & \cdots &\vdots &\vdots\\
\psi_2^{(n)}(k,l) & \psi_2^{(n+1)}(k,l) & \cdots &  
\psi_2^{(n+N-1)}(k,l)
\end{array}
\right|\,,
$$
$$
\rho_n(k,l)=
\left|
\begin{array}{cccc}
\psi_1^{(n-1)}(k,l) & \psi_1^{(n+1)}(k,l) & \cdots & 
\psi_1^{(n+N-1)}(k,l)\\
\psi_2^{(n-1)}(k,l) & \psi_2^{(n+1)}(k,l) & \cdots &  
\psi_2^{(n+N-1)}(k,l)\\
\vdots & \cdots &\vdots &\vdots\\
\psi_2^{(n-1)}(k,l) & \psi_2^{(n+1)}(k,l) & \cdots &  
\psi_2^{(n+N-1)}(k,l)
\end{array}
\right|\,,
$$
where $\psi_i^{(n)}(k,l)$ satisfies
\begin{eqnarray*}
&& \fl 
\psi_i^{(n)}(k,l)=
p_i^n(1-ap_i)^{-k}\left(1-b\frac{1}{p_i}\right)^{-l}
e^{\frac{1}{2p_i}s+\xi_{i0}-\img \pi/4}
\nonumber\\
&&\quad +(-p_i)^n(1+ap_i)^{-k}\left(1+b\frac{1}{p_i}\right)^{-l}
e^{-\frac{1}{2p_i}s+\eta_{i0}+\img \pi/4}\,,
\end{eqnarray*}
and the phase constants $\pm \img \pi/4$ play a role of 
keeping the reality and regularity. 
$s$ is an auxiliary parameter. 
Note that 
$\rho_n^m$ can be expressed as $\rho_n^m = 2\partial_s \tau_n(k,l)$ 
because the auxiliary parameter 
$s$ works on elements of the above determinant 
by $2\partial_s \psi_i^{(n)}(k,l) = \psi_i^{(n-1)}(k,l)$. 
In the lattice KdV and lattice Boussinesq equations, 
one of $\tau$-functions is 
also expressed by the derivative of another $\tau$-function 
with respect to an auxiliary parameter 
\cite{kajiwara-ohta,maruno-kajiwara}. 
This is a common property of discrete soliton 
equations which are directly connected to the B\"acklund 
transformations of continuous soliton equations.  

Let us consider Eqs.(\ref{u-k-fulldiscrete}) 
and (\ref{u-l-fulldiscrete}) again. 
Rewriting Eqs.(\ref{u-k-fulldiscrete}) 
and (\ref{u-l-fulldiscrete}), we have
\begin{eqnarray}
&&\left(\frac{u_{k+1,l}-u_{k,l}}{2}\right)^2+\delta_{k,l}^2=a^2\,,
\label{u-k-fulldiscrete2}
\\
&&\left(\frac{u_{k,l+1}+u_{k,l}}{2}\right)^2
+\left(X_{k,l+1}-X_{k,l}+\frac{1}{b}\right)^2=\frac{1}{b^2}\,.
\label{u-l-fulldiscrete2}
\end{eqnarray}
These equations actually 
give conserved quantities because $a^2$ and $1/b^2$ 
are constants. 

Introducing
\begin{eqnarray}
&&I_{k,l}\equiv
\left(\frac{u_{k+1,l}-u_{k,l}}{2}\right)^2+\delta_{k,l}^2\,,
\\
&&J_{k,l}\equiv \left(\frac{u_{k,l+1}+u_{k,l}}{2}\right)^2
+\left(X_{k,l+1}-X_{k,l}+\frac{1}{b}\right)^2\,,
\end{eqnarray}
Eqs. (\ref{u-k-fulldiscrete2}) and (\ref{u-l-fulldiscrete2}) 
imply the following conserved quantities
\begin{eqnarray}
I_{k,l}=a^2\,,\quad J_{k,l}=\frac{1}{b^2}\,,
\end{eqnarray}
for arbitrary integer values of $k$ and $l$. Hence, we have
\begin{eqnarray}
I_{k,l+1}-I_{k,l}=0\,,\quad J_{k+1,l}-J_{k,l}=0\,.
\end{eqnarray}
A substitution of the corresponding conserved quantities leads to
\begin{eqnarray}
\fl && \left(\frac{u_{k+1,l+1}+u_{k+1,l}-u_{k,l+1}-u_{k,l}}{2}\right) 
\left(\frac{u_{k+1,l+1}-u_{k+1,l}-u_{k,l+1}+u_{k,l}}{2}\right) \nonumber
\\
\fl &&\quad=-(X_{k+1,l+1}+X_{k+1,l}-X_{k,l+1}-X_{k,l})
(X_{k+1,l+1}-X_{k+1,l}-X_{k,l+1}+X_{k,l})\,,\label{discrete-eq1}
\\
\fl && \left(\frac{u_{k+1,l+1}+u_{k+1,l}+u_{k,l+1}+u_{k,l}}{2}\right)
\left(\frac{u_{k+1,l+1}+u_{k+1,l}-u_{k,l+1}-u_{k,l}}{2}\right)\nonumber
\\
\fl &&\quad=-\left(X_{k+1,l+1}-X_{k+1,l}+X_{k,l+1}-X_{k,l}+\frac{2}{b}\right)
(X_{k+1,l+1}-X_{k+1,l}-X_{k,l+1}+X_{k,l})\,.\label{discrete-eq2}
\end{eqnarray}
It can be readily shown that the difference of
Eq.(\ref{discrete-eq2}) and Eq.(\ref{discrete-eq1}) gives 
Eq.(\ref{eqX}), whereas,
the quotient is nothing but Eq.(\ref{equ}). 
In summary, Eqs.(\ref{u-k-fulldiscrete2}) 
and (\ref{u-l-fulldiscrete2}), which imply conserved quantities, 
can also be derived from 
the full-discrete analogue of the SP equation (\ref{eqX}) and
(\ref{equ}).

\section{Conclusions}
In the present paper, we proposed integrable semi-discrete and full-discrete 
analogues of the short pulse equation. The $N$-soliton solutions of both 
the continuous and discrete SP equations were formulated in the 
form of Casorati determinants, which include multi-loop soliton 
and multi-breather solutions. Based on the 
semi-discrete SP equation, a self-adaptive moving mesh method is 
proposed and used for the numerical solutions of the SP equation. 
The examples of one- and two-loop soliton solutions shows the 
potential of this novel method for the numerical study of 
the short pulse equation.

\end{document}